\documentstyle{ioplppt}
\begin{document}
\jl{4}
\title{Using inverse scattering methods to study inter-nucleus
potentials}[Using inverse scattering methods]

\author{R S Mackintosh\dag and S G Cooper\dag }

\address{\dag\ Physics Department, Open University, 
Milton Keynes MK7~6AA, UK}

\begin{abstract}
It is now straightforward to carry out $S$-matrix to potential inversion 
over a very wide
range of energies and for a wide range of projectile-target combinations.  
Inversion is possible in many cases involving  spin.
IP inversion also
 permits direct scattering data-to-potential inversion and
 furnishes powerful tools for the phenomenological
analysis of nuclear scattering. The resulting single particle potentials
exhibit various generic properties
which challenge fundamental reaction theories as well as
yield information on densities, 
provide input for reaction calculations. $S$-matrix to
potential inversion is also a powerful tool for directly investigating
theoretical processes which contribute to inter-nuclear potentials. 
Various studies have given insight into contributions to the 
dynamic polarisation potential 
(DPP) due to breakup processes and due to collective and reaction channel 
coupling and have also illuminated the role played by exchange processes in
leading to non-locality and parity dependence of the potentials. 
A case study involving d + $^4$He is a model for ways in which inversion
applied jointly to theory and experiment might illuminate
the scattering of exotic nuclei.
\end{abstract}

\maketitle
\section{Introduction}
There now exists a highly developed and indefinitely generalisable 
`inversion' technique which makes it straightforward to carry out 
$S$-matrix to potential inversion over a very wide
range of energies and for a wide range of projectile-target combinations.
This is the `iterative perturbative', IP, method. 
Inversion by means of the IP method offers many
possibilities for increasing our physical understanding of nuclear scattering
and inter-nuclear potentials, both for stable and  unstable nuclei.
Here we support this claim by: (1) briefly indicating
the underlying principles of IP inversion; (2) giving an 
overview of the range of applicability of the IP method --- this amounts to 
specifying what we mean by `inversion',  an
increasingly  general concept; (3) indicating with typical
examples the two  basic ways it can be applied, i.e.\ to theoretical
and
empirical $S_l$; and, (4) presenting a case study to show how a joint study
of theoretical and empirical $S_l$ brings potential models centre stage as
we fit data that had never been fitted and at the same time evaluate a
theoretical model for which there had been no satisfactory evaluation.
In doing this, we show the power of potential models once 
freed of historical limitations.
The case study
concerns d + $^4$He scattering. The deuteron is, of course, the archetypal
weakly bound projectile and the d + $^4$He system is one where there is 
some chance of a fully microscopic description, and also where there is a 
large amount of data. It is therefore an ideal testing ground for some of the 
physics relating to more exotic highly polarizable nuclei.

To date, most  of the applications have not been
 to nuclei far from stability; we draw attention to the
exceptions. There is considerable
potential for applications to exotic nuclei and 
there will be many  applications that we are unaware of. 
The method is implemented in a very general code IMAGO~\cite{imago}
which we are making  portable.

\section{IP inversion; range of applicability}
Historically, fixed-$l$ inversion refers to the formal determination
of the potential $V(r)$ which leads via Schr\"{o}dinger's equation to
$S_l$ for some fixed $l$ and all energies $E\rightarrow \infty$; fixed energy 
inversion is the determination of $V(r)$ from $S_l$ for all $l$ at some
fixed energy. 
Practical versions of  formal methods~\cite{chadan} for
fixed-$l$ and fixed energy inversion that can handle, for example,
finite ranges of energy or of $l$ have disadvantages 
and progress toward practical means of treating spin has been slow. Moreover,
 $S_l$ which are imprecise or  defined over small $l$ ranges are 
difficult to handle,  so 
the number of applications yielding real physical insight has been small.

A less formal method, iterative perturbative (IP) inversion, has been 
developed~\cite{early,ketal1,ketal2,candm,shirley}.
IP inversion is very general and is readily extended to include spin. It
can, moreover,
give meaningful results with $S_l$ given over small ranges of $l$ and 
also for $S_l$ which are imprecisely
determined. This last point turns out to be a crucial advantage
in many situations, particularly when experimental data is fitted.  
 
The IP method has been described a number of 
times~\cite{early,ketal1,ketal2,candm,shirley,invprob}, 
so we do no more than briefly outline the
underlying concepts. The key idea is iteratively to correct a potential
 $V(r)$ by adding terms\begin{equation}
V(r) \rightarrow V(r) + \sum c_i v_i(r) \label{basic} \end{equation}
where $v_i(r)$ are members of a suitable set of `basis functions'
and $c_i$ are amplitudes derived from linear equations arising from the
 response, assumed linear, of the elastic scattering $S$-matrix to small
changes $\delta V$ in the potential:
\begin{equation}  \delta S_l = -\frac{{\rm i} m}{\hbar^2 k}
\int_0^{\infty} (u_l(r))^2 \delta V(r) {\rm d}r. \label{integ} \end{equation}
In Equation~\ref{integ}, the radial wavefunction for angular momentum
$l$ is normalised according to $u_l(r) \rightarrow I_l(r) - S_l O_l(r)$ where
$I_l$ and $O_l$ are the incoming and outgoing Coulomb wavefunctions. The 
notation
is simplified: $V(r)$ stands for real and imaginary, central and 
spin-orbit terms which can  be expanded in different bases;
for the cases where spin and many energies  are 
treated see the papers cited above. 

IP inversion is indefinitely generalisable. The possibilities include:

\begin{description}
\item[i. Fixed energy inversion]  For
one energy and for finite set of $l$, $S_l \rightarrow V(r)$. The problem
at low energies is that there are too few partial waves to yield a usefully
defined potential. 
\item[ii. Mixed case (energy bite) inversion] 
The problem noted for fixed energy
inversion can be solved where (as is often the case) one has available the set
$S_l$ over a range of energies (`energy bite'). For a narrow bite, this is
tantamount to including ${\rm d}S_l/{\rm d}E$ as input information into 
the inversion.
\item[iii. Energy dependent inversion] Often, one expects the potential,
 particularly
the imaginary parts, to vary over the relevant energy range. In this case, IP
inversion can be extended to  determine directly an energy dependent potential.
In published cases~\cite{cmedep}, this is limited to the factored form $f(E)V(r)$ 
for each component,  but this limitation will be lifted in the future.
\item[iv. Inversion to fit bound state and resonance energies] Bound 
state and  resonance energies can be included in the input information for the 
inversion.
\item[v. Direct data to potential inversion] This new development is
discussed in Section~\ref{direct}.    
\end{description}

IP inversion can be applied to spinless projectiles, $S_l \rightarrow V(r)$, 
or spin 1/2 projectiles, $S_{lj} \rightarrow V(r) + {\bf l} \cdot
 \sigma V_{ls}(r)$, where $V$ is complex if $|S|<1$. Inversion for
spin 1 and spin 3/2 projectiles leading to vector spin-orbit potentials
has been carried out and the generalisation to tensor forces is under development.

For the case of identical bosons, one can determine potentials given 
$S_l$ for even $l$ only.
For many pairs of interacting nuclei, it is possible to determine
Wigner and Majorana terms for each  component, symbolically:
$V_{\rm W}(r) + (-1)^l V_{\rm M}(r)$. Both theory and experiment
make Majorana components obligatory in many cases. Indeed, it often emerges 
that inverting some given $S_l$ presents a choice between oscillatory pure Wigner 
potentials and relatively
smooth potentials which having Majorana components.  In some cases,  highly
oscillatory $l$-independent potentials which are $S_l$-equivalent to explicitly
parity dependent phenomenological potentials have been determined.

\section{How inversion can be applied.}
Theory and experiment both provide $S_l$, and it is often
fruitful to compare $V(r)$ inverted from
both sources for the same scattering situation.
Examples are p + $^4$He, described in Ref.~\cite{cmedep}, 
and  d + $^4$He, described in Section~\ref{rgmtheory}.
One key point is that the best basic theory (typically RGM) cannot
closely fit the data. But many qualitative features
can be extracted by inversion of  RGM $S_l$, see Section~\ref{pfrommt}.
These can  be compared with the same features extracted from
experimental data  by the use of inversion techniques, often
the {\em only\/} way of extracting these properties. 

\subsection{Analysis of theoretical $S$-matrix.}
\subsubsection{Dynamic polarization potentials} A local representation
of the dynamic polarization potential, DPP,
arising from the coupling to specific channels, follows immediately by
inverting the elastic $S_l$ given by  CC, CRC, CDCC, adiabatic model etc.\ 
and subtracting the bare potential. 
Of relevance to halo nuclei is the fact that the real part of the DPP
arising from the breakup of loosely bound projectiles tends to have~\cite{breakup}
 a characteristic
pattern of repulsion in the far surface and attraction at smaller radii. 
The geometric properties of the real potential are thereby modified and 
 so therefore are deductions concerning the nuclear size.
\subsubsection{Local equivalent potentials} Often, the most direct and efficient 
means of finding the local equivalent of 
a non-local potential is inversion. In particular,
we have determined~\cite{nonloc} the energy dependent local potential that 
represents the Perey Buck
non-local potential and verified that its energy dependence agrees well with
that derived from RGM (see below) and also the phenomenological 
energy dependence.   
\subsubsection{Potentials from microscopic theory\label{pfrommt}}
Certain theories of nuclear reactions
do not naturally yield a local potential model, yet potentials still 
provide an 
essential link to phenomenology. For {\em ab initio\/} calculations
of scattering involving light nuclei, the best existing model is probably the
resonating group model, RGM. Agreement with experiment is qualitative 
as a result of 
the necessary use of schematic  interactions and the omission of important 
configurations. Nevertheless, the fact that RGM does yield
qualitatively correct features emerges when one  determines
potentials from RGM $S_{lj}$.  Recent RGM studies
include: Ref.~\cite{npa589} which assesses the Majorana terms for helion
and alpha scattering from nuclei as heavy as $^{16}$O;  
Ref.~\cite{npa592} which determines the Majorana term for the nucleon-nucleus
potential for nuclei as heavy as $^{40}$Ca. Both papers analyse
the contribution
of specific exchange or coupling terms in the RGM kernel to the strength
and energy dependence of specific Wigner or Majorana terms in the 
inter-nucleus potential. The agreement between RGM and experiment for
p + $^4$He~\cite{cmedep} is mentioned in Section~\ref{twostep}.

\subsection{Inversion of $S$-matrix derived from experimental data}
\subsubsection{Two-step phenomenology, discrete energies} 
As an alternative to standard model-independent
optical-model phenomenology, one can fit the elastic scattering differential 
cross-section with an $S$-matrix function $S(l)$ and then invert. The second
step is by far the more straightforward. Concerning the first step, 
the good news is that phenomenology is liberated from prejudices 
concerning the form of the
potential; the bad news is the immense range of  ambiguities which emerges. 
What was originally
impossible to fit precisely becomes all too easy to fit. In fact, the 
realization of just how profound the ambiguities are, even at the level 
of essentially perfect fits, is a clear lesson which emerges. 
One must therefore judiciously reinstate
some prejudices, such as continuity with energy, and, most usefully, 
a relationship
between the behaviour of $|S(l)|$ (and to a lesser extent $\arg {S(l)}$) 
and the same
quantities derived from the static approximation Glauber model. 
This has been done for $^{16}$O + $^{16}$O~\cite{npa576}, 
$^{12}$C + $^{12}$C~\cite{npa552},
and $^{11}$Li scattering from $^{12}$C and $^{28}$Si~\cite{npa582}. 
Referring to the figures in Ref.~\cite{npa582}, one sees that it is 
indeed all too easy to
get essentially perfect fits to the data, radical ambiguities
existing even at the $S_l$ level. We argue in Ref.~\cite{npa582}
that our least unreasonable  fit
to $^{11}$Li elastic scattering data corresponds to a potential with 
a long tail in the real potential. This feature cannot be explained 
on the basis of current theories of $^{11}$Li. 
Similar potentials for $^{11}$Li + $^{12}$C were found by 
Mermaz~\cite{mermaz} 
by quite different means. There was no evidence for such a tail in
the non-halo   $^{11}$C + $^{12}$C case.
 It is widely believed that these fits reflect faulty data and that
new data will be fitted by a potential more in line with current theory. 
We keenly await the arrival of such new data, preferably for as wide 
as possible an angular
range, and for as many energies as possible. The full information
content of such data can be exploited. Since the $^{11}$Li 
work~\cite{npa582}, we have learned new ways to incorporate information 
from many energies, see Section~\ref{direct}.

\subsubsection{Two-step phenomenology, parameterised $S_{lj}(E)$\label{twostep}} 
The ambiguity problem
is much reduced when data at a series of energies is fitted with
a functional form $S_{lj}(E)$ to which one applies `mixed case'
or energy dependent inversion. Such forms of $S_{lj}(E)$ for few
nucleon systems are  typically $R$-matrix or effective range fits.
There are limits to the energy over which smooth $S_{lj}(E)$ exist,
but one can also incorporate $S_{lj}(E_i)$ for discrete energies
$E_i$ (one can also fit just the discrete $S_{lj}(E_i)$ 
blurring the distinction from the previous category.) 
For p + $^4$He scattering, a Majorana term, falling quite rapidly with 
energy, emerges, in agreement with
that found by inverting RGM $S_{lj}$. The energy
dependence of the Wigner term, largely due to knock-on exchange, is 
consistent with nucleon-nucleus phenomenology~\cite{cmedep}.

\subsubsection{Direct observable to potential inversion\label{direct}} 
Recently, a  generalisation of the IP method for direct observable to 
potential inversion has been  developed~\cite{ketal2,shirley}.
The key idea is that the linear equations by which amplitudes
$c_i$ are determined at each iteration arise from the minimisation of the
goodness of fit quantity $\chi^2$: 
\begin{equation}
\fl\frac{\partial \chi^2}{\partial c_i} 
=2\sum_{k,l}\left[\frac{\sigma_k-\sigma_k^{\rm in}}{(\Delta\sigma_k^{\rm in})^2}
\right]
\frac{\partial \sigma_k}{\partial S_l(E_k)}\frac{\partial S_l(E_k)}
{\partial c_i}
+2\sum_{n,k,l}\left[\frac{P_{kn}-P_{kn}^{\rm in}}{(\Delta P_{kn}^{\rm in})^2}
\right]
\frac{\partial P_{kn}}{\partial S_l(E_k)}\frac{\partial S_l(E_k)}
{\partial c_i} \label{new}
\end{equation}
where $\sigma_k^{\rm in}$ and $P_{kn}^{\rm in}$ are the input experimental
values of cross sections and analyzing powers respectively ($n$ indexing
the spin related observables for spin 1 systems), and
\begin{equation}
\chi^2 = \sum^N_{k=1} \left(\frac{\sigma_k-\sigma_k^{\rm in}}
{\Delta \sigma_k^{\rm in}} \right)^2 +
\sum_n \sum^M_{k=1} \left(\frac{P_{kn}-P_{kn}^{\rm in}}
{\Delta P_{kn}^{\rm in}} \right)^2. \label{fourth}
\end{equation} Since we are 
fitting data for many energies at once, the index $k$ indicates the energy 
as well as angle. The power of this approach will be evident from 
the d + $^4$He case study  described below, but we note that it has yielded 
the best phenomenological description
of p + $^{16}$O scattering to date~\cite{shirley}.

\section{Experiment and theory of d + $^4$He scattering\label{rgmtheory}}
Here we bring together RGM theory, the adiabatic model of breakup and a large 
compilation~\cite{kk} of experimental data to present an emerging picture of
d + $^4$He scattering. This is work in progress with
many issues to be settled.

\subsection{Inversion analysis of RGM calculations\label{RGM}}
The RGM $S_{lj}$ are from the multi-configuration RGM 
(MCRGM) study by Kanada {\em et al\/}~\cite{kkst},
hereafter KKST, of  elastic scattering of deuterons by $^4$He.
Up to 8 pseudo-states in the deuteron wavefunction were coupled 
to the elastic
scattering to represent S-wave breakup. A phenomenological imaginary
potential was included in order to  represent the effects of omitted open channels. The
N-N interaction was somewhat schematic in line with the demands of RGM 
calculations of this complexity. 
In particular, there was a spin-orbit but no tensor NN force. 
Both the $^4$He nucleus and the deuteron were pure S-wave 
and there was no breakup into
deuteron D-states. In Ref.~\cite{npa625} we discussed the inversion
of the KKST $S_{lj}$ for deuteron laboratory energies of 29.4 and
56 MeV  and presented complex 
potentials with spin-orbit components, plus Majorana terms for 
each component.

It follows from Figures 1 and 2 of Ref.~\cite{npa625} 
that the real, Wigner, central  part of the potentials varied in a reasonable
way with energy. At 29.4 MeV, the 
volume integral per nucleon pair was $J_{\rm R}= 499.72$ MeV fm$^3$ and 
the rms radius was 2.905  fm (c.f. Section~\ref{fitting}).
 The real Wigner spin-orbit potential was  unusual
in form, having a deep minimum at the nuclear centre, 
but was reasonable in magnitude.
It is relevant for what follows below that this shape seemed to be  well 
determined and was almost identical for both energies and 
very similar to the corresponding term determined from 
$S_{lj}$ for the
no-breakup RGM calculations of Lemere {\em et al\/}~\cite{lemere}. 
We note two provisos concerning the spin-orbit form: 
the spin-orbit potential is not determined for $r<$ the turning
 point for P-waves,  $\sim 0.5$ fm, 
so the inner cusp is just a smooth continuation of the potential further out. 
Secondly, tensor forces are omitted; while there are
no off-diagonal matrix elements generated by the RGM calculations, 
the vector potential might well be mocking up effects which are of tensor 
nature~\cite{npa625}.
Unlike the real spin-orbit terms, other components were  
different at the two energies, probably because of the marked
increase  with energy of flux into reaction channels. 

The MCRGM of KKST gave qualitative fits to experiment. The question then
arises: how well do these calculations 
describe d + $^4$He elastic scattering? It could be claimed that 
the fits are  what one might expect of any potential with correct 
overall strength and radius and a freely fitted additional imaginary term. 
We give an account below of the extent to which KKST MCRGM calculations do
lead to a potential model consistent with
good description of d + $^4$He elastic scattering.   
   
\subsection{D-wave breakup\label{breakup}}
How important is the omission of D-wave breakup in the RGM calculations?
To estimate this, we  performed adiabatic model
breakup calculations\cite{cdcc} using the code 
Adia~\cite{adia}. Inversion of $S_l$ (spin was ignored) led to 
the conclusion that as far as the real potential was concerned, 
the contribution of S-wave breakup is largely confined to
the central region, $r\le 1.5 $ fm, where it generates a considerable extra
attraction, a feature also found in RGM by Kukulin {\em et al\/}~\cite{kpsd}. 
On the other hand, D-wave breakup induces extra attraction over a 
wide radial range, but repulsion in the surface region. 
The effect is therefore to reduce the rms radius of the real potential.  

The effect on the imaginary central potential of adding D-wave breakup 
is similar to the effect on the real part. 
Instead of being confined to the centre of the
nucleus, as it is for S-wave breakup, we find added absorption
at all radii within the nucleus. However, according to
the adiabatic model, one characteristic effect
of adding deuteron breakup is the generation of a very deep 
absorptive region
near the nuclear centre. Such a deep feature is not found in the MCRGM results, 
Section~\ref{RGM}, nor, at lower energies at least, empirically, 
Section~\ref{fitting}. 
It remains to be studied
whether it is the adiabatic assumption or the neglect of exchange which
is responsible for the difference. Not only does MCRGM predict a less
absorptive region near $r=0$, but sometimes local emissivity, 
a phenomenon associated with strong non-local effects which are often
 very different from the
Perey effect. We conclude that D-state breakup is not ignorable
in d + $^4$He scattering.
The contributions to the local d + $^4$He potential 
induced by breakup  are generally consistent with
generic effects~\cite{breakup} for heavy
target nuclei. The fact that S-wave breakup induces attraction 
plus absorption only at the nuclear centre while D-wave breakup induces 
a polarization potential for all $r$ is  due in general terms to the 
fact that the intermediate state partial wave $l$ can change by 
$\pm 2$ in the latter case, but a more detailed explanation would
be welcome.  

\subsection{Fitting experimental data: d + $^4$He elastic 
scattering $\le 11.5$ MeV\label{fitting}}
There is much experimental data for d + $^4$He elastic 
scattering.
It has been carefully assembled and evaluated by Kuznetsova and 
Kukulin~\cite{kuzkuk} who also present a phase shift analysis, PSA. 
The serious problems arising with
such analyses motivate  work in preparation, see Ref.~\cite{ckmk}, 
in which  this data
is fitted in the course of demonstrating how inversion leads to a solution
of  the phase shift analysis problem. Here, our concern is 
to exploit potential models to confront experiment with RGM. Of the many  available
data sets we have fitted  two: one (set `J') comprises the angular 
distributions and vector analysing
powers of Jenny {\em et al\/}~\cite{jenny} at 5 deuteron laboratory
energies between 6.24 MeV and 10 MeV.
The second (set `SG') combines the differential cross sections at 
19 energies between 2.935 and 11.475
MeV~\cite{senhouse} and vector analysing powers at 12 energies between 
3.0 and 11.5 MeV~\cite{gruebler}, 539 data in all. 
As in Ref.~\cite{ckmk}, the differential cross sections and 
analysing powers of one of the two (J or SG) data sets for all energies are 
fitted to a single potential (which we designate respectively as the 
J or SG potential) using the direct data-to-potential form of IP 
inversion described in Section~\ref{direct}. We have not exploited the 
possibility~\cite{cmedep}
of determining a fully energy dependent potential, the real components
all being energy independent. However, the imaginary components are 
energy dependent.
When fitting the SG data, all imaginary components
are assumed to be zero below $E_{\rm th}$, $E_{\rm th}$ being the 
inelastic threshold energy. For both J and SG data, all imaginary components  
are proportional to  $(E-E_{\rm th})$ above $E_{\rm th}$. 
Tensor observables are not 
fitted and we include no representation of tensor forces, these being
understood to have a small effect on vector analysing power. 
The starting potential for the
IP procedure was that described in Section~\ref{RGM} as fitting the 29.4 MeV 
KKST $S_{lj}$. Although it is possible to include the S-wave bound 
state energy of $-1.472$ MeV in the  inversion input data, 
mostly we did not do this. Instead, the  bound state energy was
monitored to verify the consistency of the inversion.
The possible dependence of the results on the starting potential must be borne
in mind in the following discussion.

Remarkably, the inversion 
process\footnote{Each of the eight terms in the potential
had a basis of four harmonic oscillator functions, so that when SVD was 
not a limiting factor there were 32 parameters.} converged very rapidly 
to  a potential that was
rather close to the starting potential, in spite of the merely 
qualitative initial fit. Figures
1 and 2 show the fit, for  representative energies, to the SG data with
the SG potential; $\chi^2/F= 5.84$ and $E_{\rm bound} = -1.611$ MeV. 
The J potential\footnote{The fits shown 
during the oral presentation can be supplied by the author on request.}
fitted the J data equally well, $\chi^2/F= 5.95$ and $E_{\rm bound} = -1.8$ MeV.
We comment below on the values of $E_{\rm bound}$. 
In both cases, the $\chi^2/F$ are calculated using
uncertainties which do not represent all the errors discussed in the source 
references.   In Figures 3 (Wigner terms) and 4 (Majorana terms) 
we compare the starting potential, i.e. the inversion potential for KKST
MCRGM at 29.4 MeV, with the J and SG potentials.  
The (energy dependent) imaginary terms are evaluated at 8 MeV. 

The J and SG potentials  are very similar\footnote{The J potential
discussed was determined by starting the inversion from SG potential;
however, completely independent fitting led to a very similar potential.} 
and we compare them jointly with the  RGM starting point. 
For the SG potential, the real, W, central potential had
volume integral per nucleon pair $J_{\rm R}= 567.8$ MeV fm$^3$ and 
rms radius 
3.01 fm; these values for potential J were 556.5 MeV fm$^3$ and
2.86 fm. If the energy dependence of Ref.~\cite{pandp} 
applies
to $J_{\rm R}$ and to scattering from $^4$He, the value at 30 MeV lab
energy would be roughly 535 MeV fm$^3$,
rather more than the RGM value of 499.72 MeV fm$^3$, (c.f. Section~\ref{RGM}). 
The volume integral of the MCRGM real central
 W potential is thus a little less than  the J and SG data, together with
general deuteron scattering
phenomenology, suggest. While it is true that D-state breakup would
appreciably increase the depth over much of the radial range,
the adiabatic model for 40 MeV suggests little modification to $J_{\rm R}$ 
due to the influence of repulsion in the far surface on the $r^2$ weighted
integral for $J_{\rm R}$. 
Nevertheless, overall  the RGM and empirical 
potentials are similar in shape.
The most striking  difference lies in the W, central,
imaginary part; this is  reasonable since 
 the flux into inelastic channels must increase 
markedly between 8 and 29.4 MeV. The emissive
region at the nuclear centre for this imaginary term is firmly established
by inversion. This is not unusual in RGM inversion studies and there is
 a nearly emissive feature  at the centre of the MCRGM
potential,  probably due
to non-locality. There is nothing comparable in the unsymmetrised
adiabatic breakup calculations.
The W spin-orbit terms broadly follow the RGM shape. We have pointed
out that the 
spin-orbit potential is not determined for $r < 0.5$ fm 
and  the central cusps are probably an artifact of the harmonic oscillator
basis. One curious feature is that fact that potential J has 
$J_{\rm R}$ which is slightly smaller in magnitude than that of potential SG. 
This is, of course, consistent with the higher mean energy of the 
data to which
it is fitted, but the associated deeper binding for 
potential J is somewhat counter-intuitive. From another perspective it is
not, perhaps, surprising that the potential (SG)
which is constrained by data reaching to lower energies gives a better 
fit to
the experimental binding energy, $-1.472$ MeV. However, 
the calculation of $J_{\rm R}$ and, even more the rms radius, demands 
that the potential be very well determined in the far surface; we cannot
absolutely guarantee this at present.

Concerning the much smaller M terms in Figure 4:
unsurprisingly,  the J and SG potentials jointly differ  more 
from the 29.4 MeV starting potential, and are probably poorly determined
for $r< 1$ fm. 
The attractive real central term for $r>1.5 $ fm, where J and SG 
agree with RGM, seems to be well established. 

These studies are continuing; the energy range will be increased, and 
the possibility that some irregular features in the potential are
artefacts of an inadequate energy dependence and/or lack of
tensor interaction studied.

\section{Conclusions: what IP might do for the study of halo nuclei}
Observable-to-potential IP inversion 
will find a local potential to fit any elastic scattering data. 
If data for several energies are available,
so much the better, since continuity with energy is a useful criterion 
in evaluating ambiguities when what was otherwise 
impossible to fit precisely becomes all too easy to fit. 
In some cases, other {\em a priori\/} information can be incorporated.

IP inversion can contribute to a theoretical understanding of 
 potentials. It readily yields local potentials
corresponding to any theoretical model whatever; thus one can 
relate modifications to the model to changes induced
in the potential. This opens the way to finding a rich variety of 
generic properties of  the DPP or exchange contributions 
(especially for RGM), pointing to what is required to make 
good the difference between theoretical and empirical potentials. 

Inversion for both experiment and theory 
for the same reaction makes local potentials  of special value. 
The p + $^{4}$He case
discussed elsewhere, and the d + $^4$He case discussed here show 
this. We have concluded that the KKST model 
gives a very good description of  d + $^4$He scattering even though
 the fits presented by KKST seem mediocre.
For example, theory and data substantially agree the on the nature
of the real Majorana terms; the imaginary Majorana terms probably require 
D-state breakup which we know is required in a complete description. 
A host of interesting theoretical puzzles arise from this work:
for example, just how do non-locality effects lead to
an emissive region at the nuclear centre rather than the deep 
absorptive region
firmly predicated by unsymmetrised breakup calculations?  

\ack
We are grateful to the EPSRC for grant GR/H00895 supporting Dr Cooper. We are
particularly indebted to Professor V.I. Kukulin of Moscow for
providing numerical data and much inspiration.

\Figures

\Figure{For deuterons scattering from $^4$He, 
the differential cross-sections calculated with potential SG
compared with data measured at 6 energies
in the range from 2.93 MeV and 11.475 MeV as measured by
 Senhouse and Tombrello. Potential SG was determined as described
in the text by fitting differential cross-sections for all 19 energies and 
analysing power for 12 energies.}

\Figure{For deuterons scattering from $^4$He
 the vector analysing powers calculated with potential SG
compared with data measured  at 6 energies
in the range from 3.00 MeV and 11.0 MeV measured by
Gruebler {\em et al\/}. Potential SG was determined as described
in the text by fitting analysing powers for all 12 energies and cross-sections
 for 19 energies.}

\Figure{The Wigner (W) components of the starting potential (fitting KKST 
$S_{lj}$) and potentials J and SG. From the top we present the real and
imaginary central, then real and imaginary spin-orbit components.
The solid line is the KKST fit for 29.4 MeV lab, the dashed line is
potential SG and the dotted line is potential J.}

\Figure{The Majorana (M) components of the same potentials as in the
previous figure, with the same order and same conventions.}

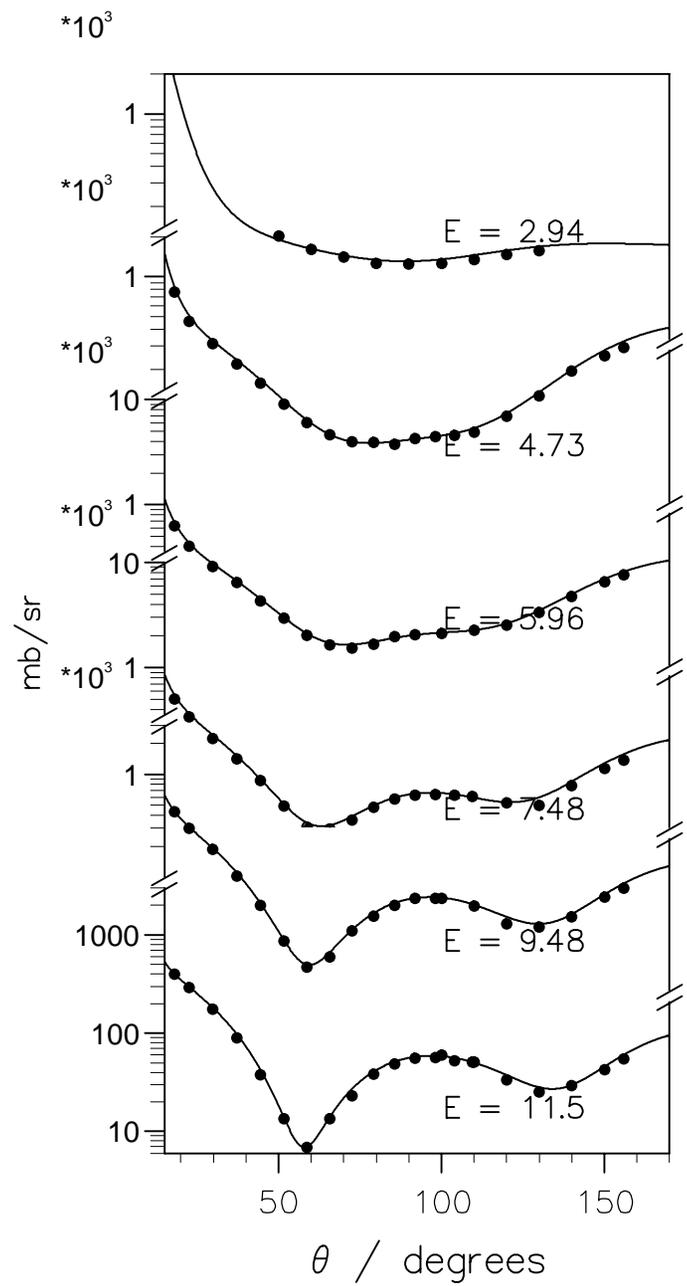
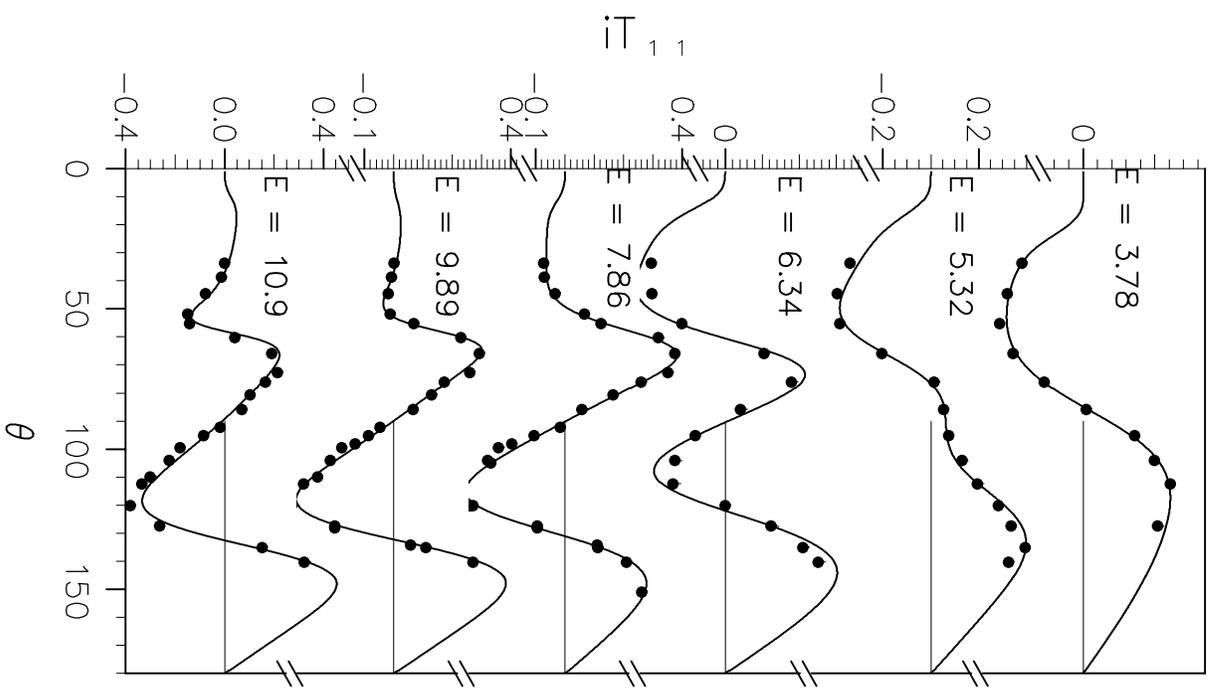
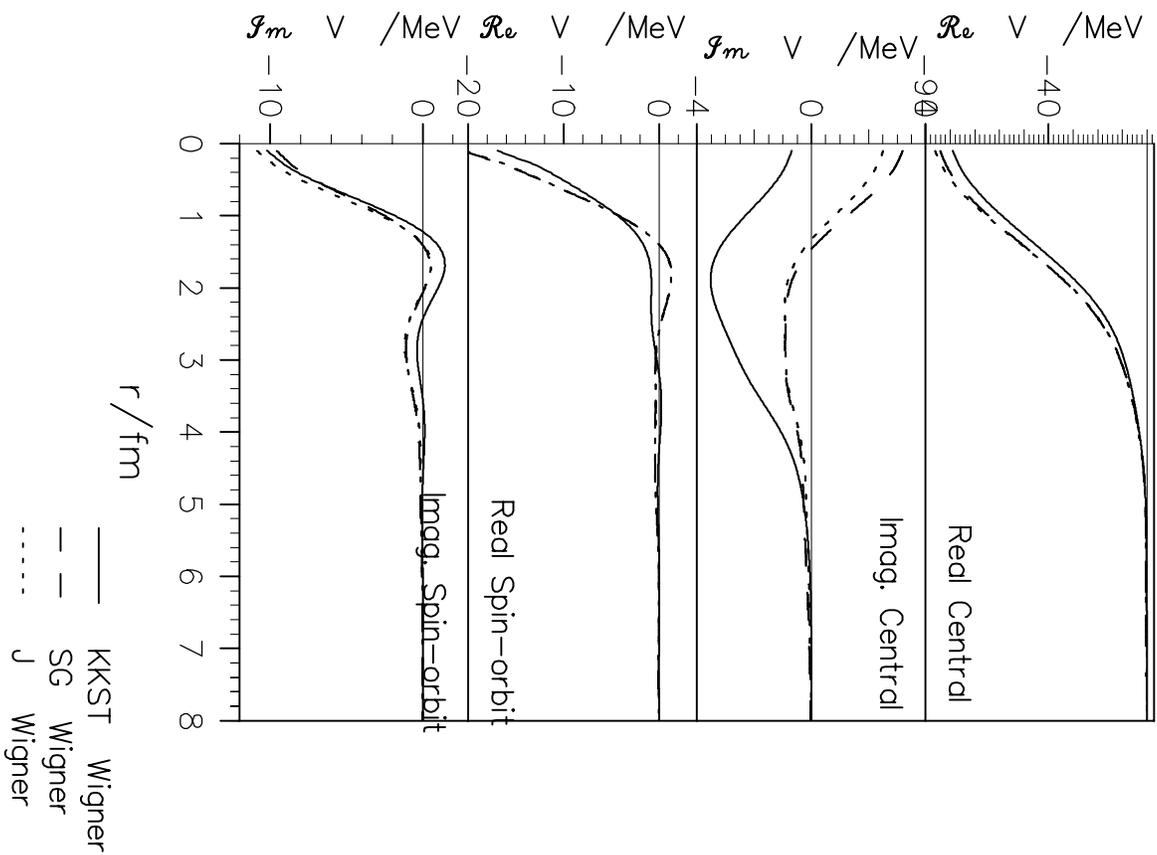
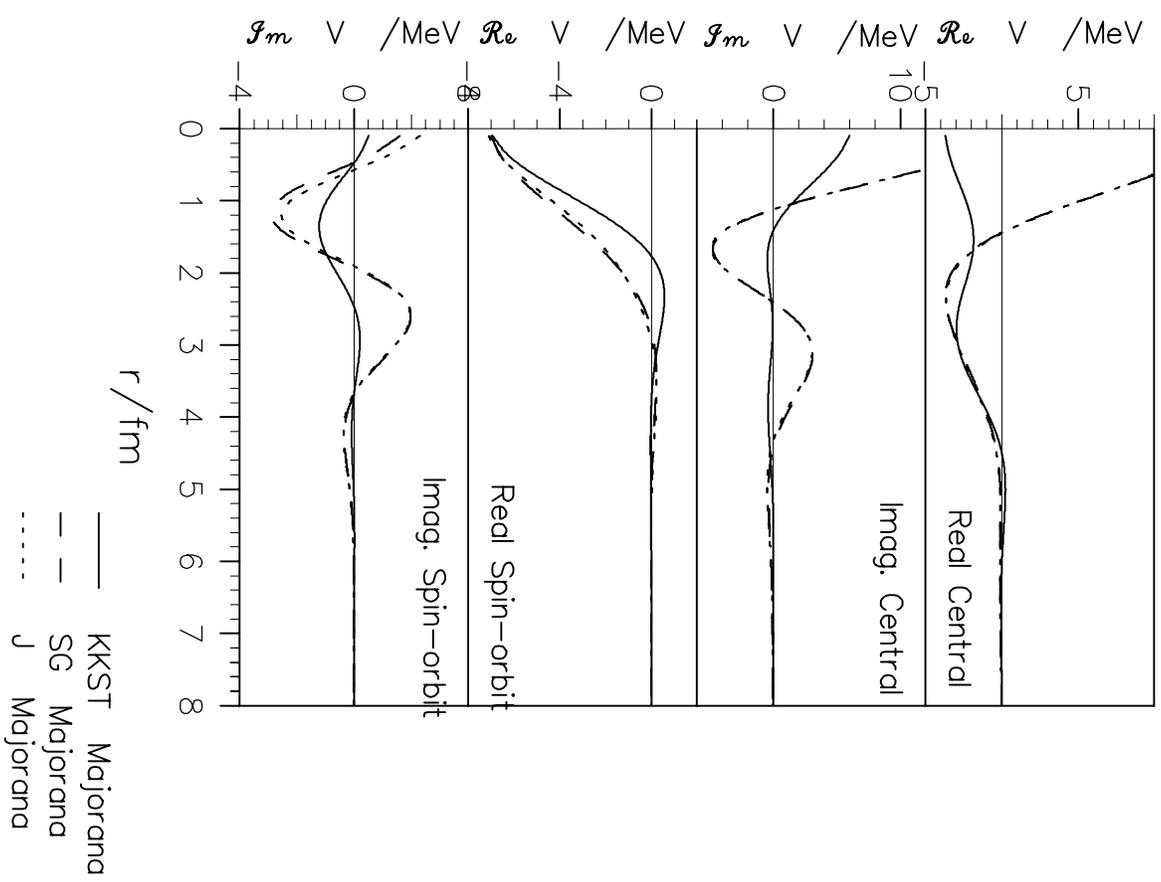
\end{document}